\documentclass[a4paper]{jpconf}
\usepackage{graphicx}
\begin{document}
\title{On the microwave response of thin superconducting films with trapped magnetic flux}

\author{D A Luzhbin}

\address{Superconductivity department, Institute for Metal Physics, 03142 Kyiv,
Ukraine}

\ead{luzhbin@imp.kiev.ua}

\begin{abstract}
Basing on the phenomenological model of the microwave response of
thin superconducting films with trapped magnetic flux elaborated in
Luzhbin {\it et al} 2005 \SUST {\bf 18} 1112-7, detailed analysis of
the ac linear response of a superconducting strip-line resonator
cooled down in FC and ZFC regimes is presented. Special emphasis on
a nonuniform distribution of the trapped vortices and on the
magnetic field and temperature dependence of the phenomenological
model parameters such as critical current density, vortex pinning
and viscous motion coefficients is made. For a ZFC case realistic
trapped magnetic field distribution based on the critical-state
model for a vortex penetration is used. Within the framework of the
used model the unusual multi-peak dependence of the surface
resistance of HTS films observed experimentally is explained as well
as clarified conditions for its appearance.
\end{abstract}

\section{Introduction}

The phenomenological theory of the linear response of type-II
superconductors (SC) in the mixed state developed in the last
decades \cite{th1,th2,th3,th4,th5} describes the experimental data
sufficiently well. Being based mainly on the idea of the induced
current and its effect on the quasi-particles due to vortex motion,
it can be formulated in terms of a change of the local effective
resistivity of the superconducting material. Introducing into the
model description various phenomenological parameters (e.g. pinning
constant $\alpha_L$, viscosity $\eta$, creep rates $\tau$, $\tau_0$
\cite{th2,th3}, critical current density $J_c$, elastic moduli of
the flux line lattice $C_{ii}$, $i=1,4,6$, etc) allows us to
simultaneously incorporate into the theory various effects of the
external magnetic field and the temperature starting from the
ordinary flux creep to the phase transitions in the vortex ensemble.

However, a number of experiments shows unusual quasi-periodic
dependence of the surface impedance of HTS films on the temperature
or the external dc magnetic field \cite{ex4,ex5,ex6}, which cannot
be unambiguously explained within the frameworks of the existing
theoretical models.

In the current report the theoretical model of the linear response
of HTS thin films with trapped vortices developed in \cite{my} is
used to model the dc magnetic field and the temperature dependence
of energy losses $P_{\rm dis}(B,T)$ connected with the oscillation
of Abrikosov vortices in thin SC films cooled down in (FC) and
without (ZFC) external dc magnetic field. It is shown that, under
the general assumptions on the magnetic field and temperature
behavior of $J_c$, $\alpha_L$ and $\eta$, the quasi-periodic form of
$P_{\rm dis}(T)$ dependence neatly corresponds to the ones obtained
in \cite{ex4,ex5,ex6}. By varying the shape of $J_c(B,T)$,
$\alpha_L(B,T)$ and $\eta(B,T)$ dependences, the conditions for the
appearance of nonmonotonic parts on $P_{\rm dis}(B,T)$ curve is also
clarified.

\section{Theoretical consideration}

Let us consider a thin SC film (with thickness $d \le \lambda$ and
width $2w \gg 2\lambda^2/d$) with a trapped magnetic flux. It is
supposed that all physical quantities (current, magnetic field, etc)
are distributed along the $y$-axis only and averaged over the film
thickness, i.e. $J(y)=\int_{-d/2}^{d/2}J(y,z) dz/d$.

The trapped magnetic flux is produced by Abrikosov vortices
distributed along the film width with some magnetic field profile
$B(y)$, which is proportional to the local density of vortices. The
profile $B(y)$ in turn depends on the cooling prehistory of the film
and can in principle be found by means of magneto-optical, scanning
Hall probe microscopy, or Lorentz microscopy methods.

Upon the application of an external ac current $J_{\rm
rf}(y)\exp(-i\omega t)$, vortices are forced to oscillate around the
equilibrium positions with the same frequency $\omega$. The vortex
displacement field $u(y)$ can be found from the force balance
condition \cite{th2,th3}, which in our case has the form \cite{my}:
\begin{equation} 
J_{\rm rf}(y)-\frac{2}{\pi\mu_0d}\int_{-w}^{w}\frac{[B(\zeta)u(\zeta)]'}{\zeta-y}\sqrt{\frac{w^2-\zeta^2}{w^2-y^2}}\,d\zeta
 =\frac{\widetilde{\alpha}}{\Phi_0}u(y) \,, \label{e1}
\end{equation}
where $\widetilde{\alpha}=\alpha_L(1-\rmi\omega/\omega_0)$,
$\omega_0=\alpha_L/\eta$ is the depinning frequency that lies in the
range $10^{10} - 10^{12}$~s$^{-1}$ \cite{golos}, and $\Phi_0$ is the
flux quantum. Equation \eref{e1} models the well known ac current
concentration in superconducting strip-line resonators at the film
edges by a function with the square root singularities at the edges
of the film, which is a rather good approximation \cite{cur,skoryk}.
Moreover, since equation \eref{e1} has the finite solution over the
whole interval $-w \le y\le w$, it proposes a very realistic model
of the linear response of thin SC films with trapped vortices.

Having found the vortex displacement field $u(y)$ as the solution to
\eref{e1}, one can calculate the response of the film to the ac
current; in particular, the ohmic losses $P_{\rm dis}$ per unit
length of the film during the period of the ac field oscillations
can be found according to the definition \cite{norris}
\begin{equation} 
P_{\rm dis}=\frac{gd}{2} {\rm Re}\left (\int_{-y_{\rm edge}}^{y_{\rm
edge}}J(y,t)E^*(y,t)\,dy \right )+P_{\rm edge}\,, \label{e2}
\end{equation}
where all time dependent variables are supposed to have the form
$\exp(-i\omega t)$, $g$ is defined by the geometry and operation
regime of the resonator, $y_{\rm edge}=w-2\lambda^2/d$, and $P_{\rm
edge}$ corresponds to the ohmic losses at the edges of the film,
i.e. for $y_{\rm edge} \le |y| \le w$, $J(y)$ is the total current
in the resonator, and $E(y)$ is the ac electrical field generated by
the current $J(y)$. In \eref{e2} the fact that the real current
distribution in the resonator is obtained by the current cut off at
a distance of order $2\lambda^2/d$ from the edges \cite{skoryk} is
taken into account.

The general idea of the numerical solution of equation \eref{e1}
that gives the approximate solution with an arbitrary specified
precision can be summarized as follows \cite{AML}.
\begin{enumerate}
\item
For any smooth function B(y) that has no zeros in the interval
$[-w,w]$, the function $f(y)=B(y)u(y)$ can be well approximated by a
polynomial of degree $N$, given by $f(y)\simeq
\sum_{j=0}^{N}a_j\widetilde{y}^j$, where $a_j$ are the unknown
complex coefficients, $\widetilde{y}=y/w$, and $f'(y)\simeq
\sum_{j=0}^{N}ja_j\widetilde{y}^{j-1}$.
\item
Let us write $B(y)=B_{\rm dc}f_B(\widetilde{y})$ and $J_{\rm
rf}(y)=J_{\rm rf,0}f_J(\widetilde{y})$, where $B_{\rm dc}$ and
$J_{\rm rf,0}$ are some constants that have dimensions of magnetic
field and current density, respectively, and the functions $f_B$ and
$f_J$ define the dc magnetic field profile and the ac current
distribution over the film width, respectively. In general,
$f_B(\widetilde{y})$ also depends on the external dc field
\cite{BrInd}, but in the simplest case of the uniform vortex
distribution $f_B$ is a function of $\widetilde{y}$ only, and
$B_{\rm dc}$ has the meaning of the external dc field (e.g. for FC
case). For the linear response it is always possible to write $a_j =
dw\mu_0J_{\rm rf,0}c_j$, where $c_j$ are dimensionless unknown
coefficients. Multiplying both sides of equation \eref{e1} by $B(y)$
yields the equivalent equation
\begin{equation} 
f_J(\widetilde{y}) f_B(\widetilde{y})-\frac{2
f_B(\widetilde{y})}{\pi\sqrt{1- \widetilde{y}^2}}\sum_{j=1}^N
jc_j\varphi_j(\widetilde{y})=S(\omega)\sum_{j=0}^N c_j
\widetilde{y}^j\,, \label{e3} \end{equation}
where $S(\omega) = dw/\lambda_C^2(\omega)$,
$\lambda_C^2(\omega)=B_{\rm dc}\Phi_0/\mu_0\widetilde{\alpha}$ is
the Campbell penetration depth \cite{th2,th3}, and
$\varphi_j(\widetilde{y})=\int_{-1}^1
\zeta^j\sqrt{1-\zeta^2}\,d\zeta\,/(\zeta-\widetilde{y})$.
\item
Taking the values of all functions in \eref{e3} at the basic points
$\widetilde{y}_0$, $\widetilde{y}_1$, ... $\widetilde{y}_N$ which
are the zeros of the Chebyshev polynomial
$T_{N+1}(\widetilde{y})=\cos((N+1)\arccos(\widetilde{y}))$, one
obtains a system of linear equations for determining the unknown
coefficients $c _j$.
\end{enumerate}

Since the parameters defining the response of resonators in this
approximation (i.e. $f_J(\widetilde{y})$ and $f_B(\widetilde{y})$)
are supposed to be known from elsewhere, this model is directly
applicable to any type of thin film resonator. The quasi-periodic
form of $P_{\rm dis}(S(\omega))$ dependence that follows from
equations \eref{e1}, \eref{e3} (see \fref{pic1}(A)) is a
manifestation of the dimensional effect associated with quantization
of the displacement field $u(y)$ on the width of the film \cite{my}.

To model the temperature or the dc magnetic field dependence of the
microwave response in the framework of this model, one should take
into account the following factors: the temperature/magnetic field
dependence of the $f_B$ profile (e.g. due to flux creep), and the dc
magnetic field and temperature dependence of the phenomenological
parameters $\alpha_L$, $\eta$ and $J_c$, which is supposed to be
known from elsewhere, see, e.g. \cite{golos, ghosh,myprb}.

For the sake of an example let us consider the temperature
dependence of the energy losses for a film cooled down in FC and ZFC
regimes in an uniform external magnetic field $B_{\rm dc}$. For the
FC case the uniform distribution of the trapped vortices is a rather
good fit, i.e. $B(y)=B_{\rm dc}$, whereas for the ZFC case the
critical state model \cite{BrInd} is the best approximation. The
original critical state model is modified in the following way:
$B(y)=B_{\rm remn}$ for $|y|<b$, and $|J(y)|=J_c$ for $b<|y|<w$ with
$b=w/\cosh(\pi B_{\rm dc}/\mu_0J_c)$ and $B_{\rm remn}\ll B_{\rm
dc}$ being some remnant field. This modification allows us to use
the foregoing approach to model the ac response of the film in ZFC
state and, as long as $B_{\rm remn}\ll B_{\rm dc}$, it does not
affect the results; on the other hand, such modification is
justified by the fact that in a lot of experiments some unscreened
magnetic field of order of several oersted is always present
\cite{ex6}. In both cases, the following temperature \cite{golos}
and the dc magnetic field \cite{ghosh} dependence of $\alpha_L$ and
$\eta$ is supposed:
\begin{equation} 
\eqalign{\alpha_L(T)=\alpha_L(0)(1-t)^{4/3}(1+t)^2\exp(-T/T_0)\,,~\alpha_L(B)\propto B_{dc}^{n_\alpha}\,,\\
\eta(T)=\eta(0)(1-t^2)/(1+t^2)\,,~\eta(B)\propto B_{dc}^{n_\eta}\,,}
\label{e4}
\end{equation}
where $t=T/T_c$, $T_c=88K$, typical values of $\alpha_L(0)\simeq 3
\times 10^5$~N~m$^{-2}$ and $\eta(0) \simeq (0.2-1.2) \times
10^{-6}$~N~s~m$^{-2}$, $T_0$ is a characteristic fluctuation
temperature \cite{th5,golos} (with $T_0\simeq 28K$ for YBCO),
$n_{\alpha}\simeq 0.55$, $n_{\eta}\simeq 1.85$ \cite{ghosh}. Note
that the given values of $\alpha_L(0)$, $\eta(0)$ as well as their
temperature and field dependences \eref{e4} are the best fit for the
numerous experimental data obtained in a number of experiments on
YBCO \cite{golos, ghosh}, whereas theoretical predictions vary from
the field-independent $\eta$ for the isotropic s-wave pairing to
$n_{\eta}=1/2$ for the d-wave pairing, and from the
field-independent $\alpha_L$ for the individual pinning of vortices
to $n_{\alpha}=1$ for the collective pinning.

The temperature and the magnetic field dependence of $J_c$ is well
described by the following empirical formula \cite{myprb}:
$J_c(B,T)=J_{c0}(B)(1-t)^{\varepsilon}$, where $J_{c0}(B)=J_{c0}(0)
\beta \ln(B^{\star}/B)$ in the range of interest, $B^{\star} \simeq
B_{\rm eff}(1-t)$. $J_{c0}(0)$, $\varepsilon$, $\beta$ and $B_{\rm
eff}$ are some empirical parameters, which can be independently
obtained from transport, ac susceptibility and magnetization
measurements. For YBCO films, the following values are typical:
$\beta$, $\varepsilon$ are of order of unity, $B_{\rm eff}$ of order
of several tesla, and $J_{c0}(0)$ of order of several MA~cm$^{-2}$
\cite{myprb}.

\section{Discussion}

\begin{figure}[h]
\begin{minipage}{18pc}
\includegraphics[width=18pc]{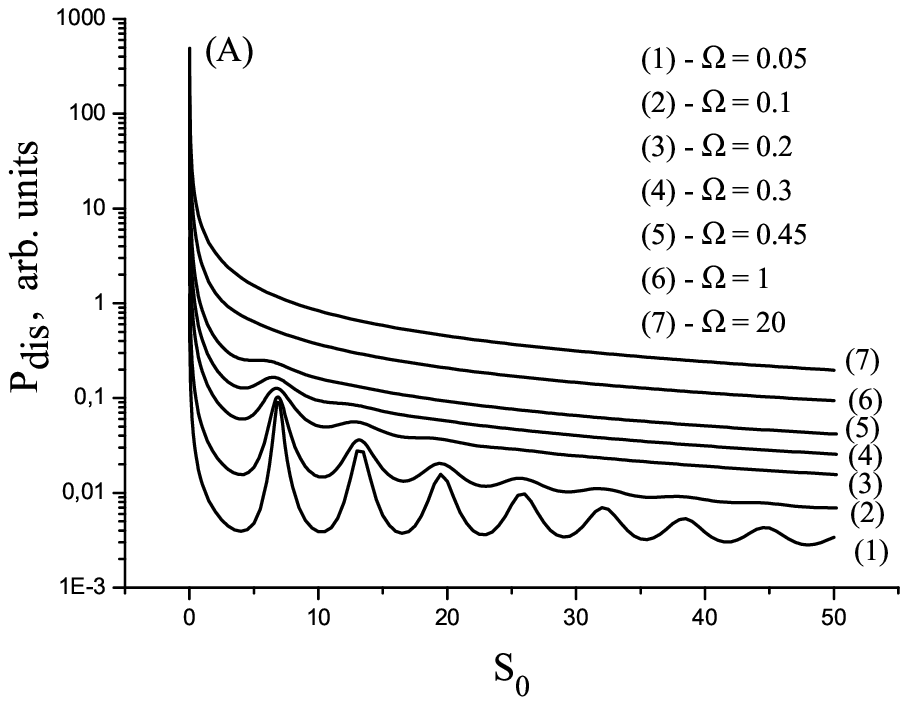}
\end{minipage}\hspace{2pc}%
\begin{minipage}{18pc}
\includegraphics[width=18pc]{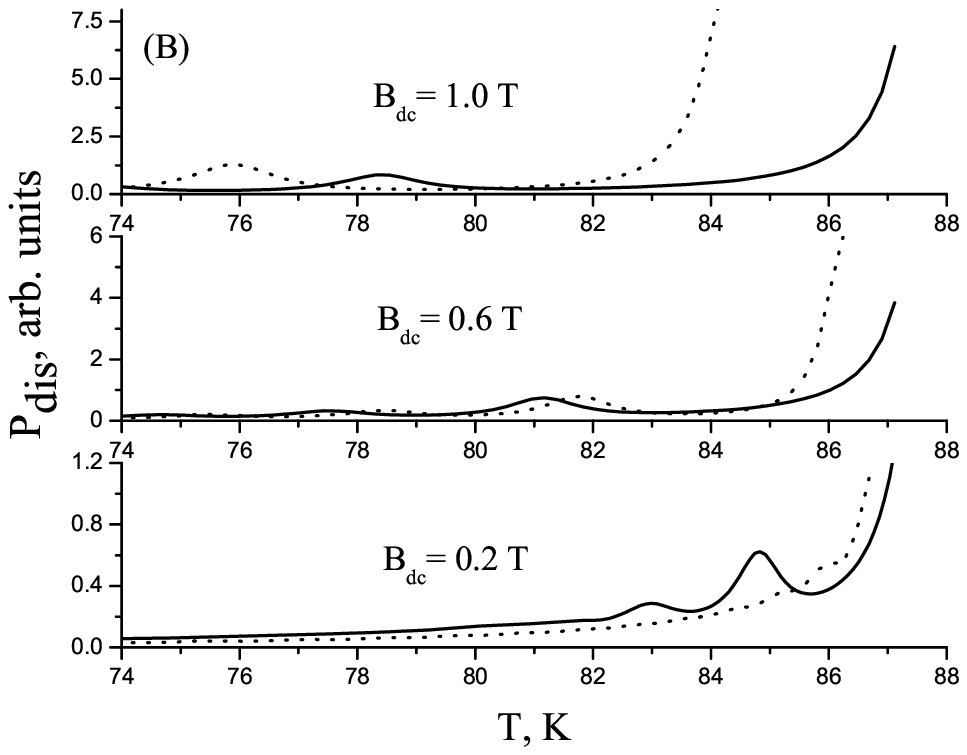}
\end{minipage}
\caption{\label{pic1}Vortex contribution to the ohmic losses in the
film (A) as a function of the parameter
$S_0=dw/\lambda_C^2(\omega=0)$, $\Omega=\omega/\omega_0$, and (B) as
a function of the temperature for FC (\full) and ZFC (\dotted)
regimes for the frequency $\omega/2\pi=1$~GHz. To avoid crowding,
ZFC curves are multiplied by factor 4.}
\end{figure}

The energy loss dependence on the temperature $P_{\rm dis}(T)$ for a
film in FC and ZFC states is shown in \fref{pic1}(B). All the
parameters $\alpha_L$, $\eta$, $\omega_0$ and $J_c$ are supposed to
be temperature and magnetic field dependent according to \eref{e4}
with $\alpha_L(0)=10^5$~N~m$^{-2}$ and $\eta(0)=0.2\times
10^{-6}$~N~s~m$^{-2}$.

It is found that the exact form of $P_{\rm dis}(B,T)$ curve and the
peak positions are mainly defined by $\alpha_L(0)$ and $J_{c0}(0)$
as well as the exact form of $J_c(B,T)$ dependence, while the
quantities $\eta$ and $\omega_0$ define the peak amplitudes.
Decrease of $\alpha_L(0)$ and $J_{c0}(0)$ shifts the position of the
peaks to the lower temperatures and decreases their amplitude, so
that the peaks cannot be detected.

Other crucial factors are the geometrical parameters of a resonator,
i.e. its thickness $d$ and width $2w$. As it is seen from
\fref{pic1}(A), the most pronounced peaks in the curve $P_{\rm
dis}(S_0)$ are located in the region of tens of $S_0$; thus, only if
the dimensions of a resonator lie in the range corresponding to $5
\le S_0 \le (20-30)$ for the rather low frequency, the multi-peak
picture similar to the ones obtained in \cite{ex4,ex5,ex6} can be
experimentally observed.

In that way, the possibility to experimentally observe the
quasi-periodic (multi-peak) dependence $P_{\rm dis}(T)$ crucially
depends on the transport properties of HTS films, which can be
characterized by $J_c$ and $\alpha_L$ and in turn are determined by
the conditions and method of preparation of the films, as well as on
the geometrical dimensions of SC resonators.

\ack

The work is supported by the NATO Reintegration Grant
NUKR.RIG.981505

\end{document}